\documentclass[]{spie}  

\usepackage{amsmath,amsfonts,amssymb}
\usepackage{graphicx}
\usepackage{footmisc}
\usepackage[colorlinks=true, allcolors=blue]{hyperref}

\title{CCAT-prime: Design of the Mod-Cam receiver and 280~GHz MKID instrument module}

\author[a]{Eve M. Vavagiakis}
\author[a]{Cody J. Duell}
\author[b]{Jason Austermann}
\author[b]{James Beall}
\author[c]{Tanay Bhandarkar}
\author[d]{Scott C. Chapman}
\author[a,g]{Steve K. Choi}
\author[e,f]{Gabriele Coppi}
\author[c]{Simon Dicker}
\author[c]{Mark Devlin}
\author[g]{Rodrigo G. Freundt}
\author[b]{Jiansong Gao}
\author[h]{Christopher Groppi}
\author[g]{Terry L. Herter}
\author[a]{Zachary B. Huber}
\author[b]{Johannes Hubmayr}
\author[i,j]{Doug Johnstone}
\author[a]{Ben Keller}
\author[c]{Anna M. Kofman}
\author[a,k]{Yaqiong Li}
\author[h]{Philip Mauskopf}
\author[l]{Jeff McMahon}
\author[h]{Jenna Moore}
\author[a]{Colin C. Murphy}
\author[a,g]{Michael D. Niemack}
\author[m]{Thomas Nikola}
\author[c]{John Orlowski-Scherer}
\author[m]{Kayla M. Rossi}
\author[n]{Adrian K. Sinclair}
\author[g]{Gordon J. Stacey}
\author[b]{Joel Ullom}
\author[b]{Michael Vissers}
\author[b]{Jordan Wheeler}
\author[o]{Zhilei Xu}
\author[c]{Ningfeng Zhu}
\author[p]{Bugao Zou}

\affil[a]{Department of Physics, Cornell University, Ithaca, NY, 14853, USA}
\affil[b]{Quantum Sensors Group, NIST, Boulder, CO, 80305, USA}
\affil[c]{Department of Physics and Astronomy, University of Pennsylvania, Philadelphia, PA 19104, USA}
\affil[d]{Department of Physics and Atmospheric Science, Dalhousie University, Halifax, NS, B3H 4R2, Canada}
\affil[e]{Department of Physics, University of Milano-Bicocca, Piazza della Scienza 3, 20126 Milano, Italy}
\affil[f]{National Institute for Nuclear Physics (INFN), Sezione di Milano-Bicocca, Piazza della Scienza 3, 20126 Milano, Italy}
\affil[g]{Department of Astronomy, Cornell University, Ithaca, NY, 14853, USA}
\affil[h]{School of Earth and Space Exploration, Arizona State University, Tempe, AZ, 85281, USA}
\affil[i]{National Research Council, Herzberg Astronomy and Astrophysics, Victoria, BC, V9E 2E7,Canada}
\affil[j]{Department of Physics and Astronomy, University of Victoria, Victoria, BC, V8P 5C2, Canada}
\affil[k]{Kavli Institute at Cornell for Nanoscale Science, Cornell University, Ithaca, NY, 14853, USA}
\affil[l]{Department of Astronomy and Astrophysics, University of Chicago, Chicago, IL, 60637, USA}
\affil[m]{Cornell Center for Astrophysics and Planetary Sciences, Cornell University, Ithaca, NY,
14853, USA}
\affil[n]{Department of Physics and Astronomy, University of British Columbia, Vancouver, BC, V6T 1Z1, Canada}
\affil[o]{MIT Kavli Institute, Massachusetts Institute of Technology, 77 Massachusetts Avenue, Cambridge, MA 02139, USA}
\affil[p]{Department of Applied and Engineering Physics, Cornell University, Ithaca, NY 14853, USA}

\authorinfo{Further author information: (Send correspondence to E.M.V.)\\E.M.V.: E-mail: ev66@cornell.edu, Telephone: 1 607 255 0474}

\begin{document} 
\maketitle

\begin{abstract}
Mod-Cam is a first light and commissioning instrument for the CCAT-prime project’s six-meter aperture Fred Young Submillimeter Telescope (FYST), currently under construction at 5600 m on Cerro Chajnantor in Chile’s Atacama Desert. Prime-Cam, a first-generation science instrument for FYST, will deliver over ten times greater mapping speed than current and near-term facilities for unprecedented 280--850 GHz broadband and spectroscopic measurements with microwave kinetic inductance detectors (MKIDs). CCAT-prime will address a suite of science goals, from Big Bang cosmology to star formation and galaxy evolution over cosmic time. Mod-Cam deployment on FYST with a 280~GHz instrument module containing MKID arrays is planned for early science observations in 2024. Mod-Cam will be used to test instrument modules for Prime-Cam, which can house up to seven instrument modules. We discuss the design and status of the 0.9 m diameter, 1.8 m long Mod-Cam receiver and 40 cm diameter 280~GHz instrument module, with cold stages at 40~K, 4~K, 1~K, and 100~mK. We also describe the instrument module’s cryogenic readout designs to enable the readout of more than 10,000 MKIDs across 18 networks.  
\end{abstract}


\section{INTRODUCTION}
\label{sec:intro}  

\begin{figure}[h!]
\centering
    \includegraphics[width=1.0\linewidth]{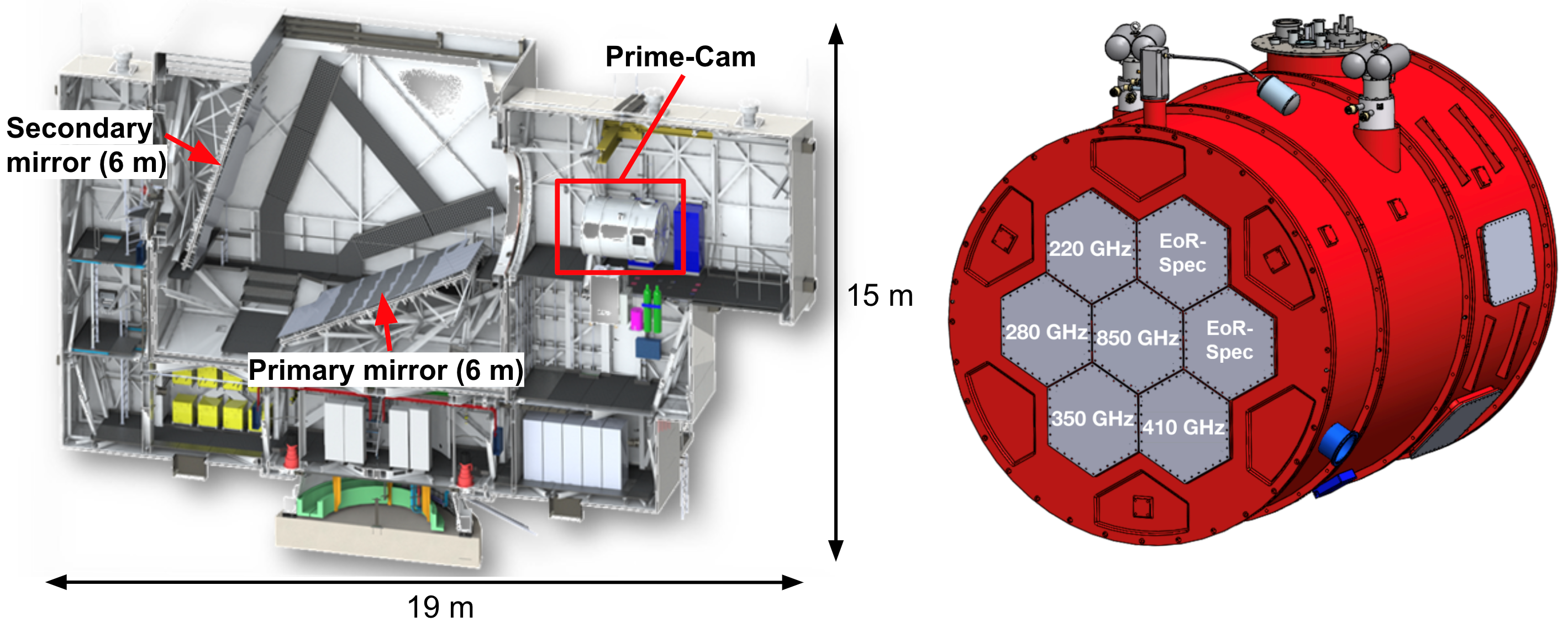}
    \caption{Left: A cross section of the FYST model, revealing the 6 meter primary and secondary mirrors which focus light into the instrument space where Prime-Cam or Mod-Cam will be installed (Prime-Cam model shown in render). Right: A model of Prime-Cam with a possible instrument module configuration (Figure from Ref.~\citenum{ChoiPrimeCam}).}
    \label{fig:fystpcam}
\end{figure}

The Fred Young Submillimeter Telescope (FYST) will be a 6-meter aperture crossed-Dragone telescope for the CCAT-prime Observatory at 5600~m elevation on Cerro Chajnantor in the Atacama Desert, Chile \cite{CCATscience,Parshley2022}. Prime-Cam, a 1.8-meter diameter cryogenic receiver, will be a first-generation science instrument for FYST, taking advantage of the high-efficiency telescope and high-elevation site to enable wide-field and deep mapping between 100 and 900 GHz (Fig.~\ref{fig:fystpcam}) \cite{EMVSPIE2018,ChoiPrimeCam}. FYST is currently under construction by CPI Vertex Antennentechnik GmbH in Germany, and
will begin assembly at the Cerro Chajnantor site for first light in 2024. The Simons Observatory (SO)\cite{SO:2019} Large Aperture Telescope\cite{Xu2021} shares most aspects of the mechanical design and the same optical design as FYST, making it natural for the Prime-Cam and instrument module designs to evolve from the design of the SO Large Aperture Telescope Receiver\cite{Zhu2021} and modules.

Prime-Cam's millimeter and sub-millimeter measurements will enable new science goals as well as overlap with surveys at other frequencies for synergistic analyses. Prime-Cam will house up to seven independently developed $\sim$41 cm diameter instrument modules, each with up to 1.3$^{\circ}$ field of view, filling a total of 4.9$^{\circ}$ of FYST's 8$^{\circ}$ diameter field of view at 3 mm. When fully populated, Prime-Cam will field up to five broadband polarization-sensitive modules for observations between 220 and 850 GHz, and at least two imaging spectrometer modules for line intensity mapping from 210 to 420 GHz. \cite{ChoiPrimeCam}. When populated with seven instrument modules, each deploying three microwave kinetic inductance detector (MKID) arrays, Prime-Cam will field a total of $>$100,000 detectors, larger than any deployment of broadband KIDs yet. Together, these modules will target science goals ranging from Big Bang cosmology through reionization and the formation of the first galaxies to energetic transients, galaxy cluster evolution via the Sunyaev-Zel’dovich (SZ) effects, galactic polarization and star formation within the Milky Way \cite{CCATscience}. Prime-Cam is currently under construction by Redline Chambers for delivery to Cornell University for initial testing in 2022. 

Mod-Cam is a single instrument module cryogenic receiver for Prime-Cam and a first light instrument for FYST, currently in testing at Cornell University \cite{DuellSPIE}. The first 280~GHz MKID array for the CCAT-prime project is currently in testing at Cornell University\cite{Choi2021}, and Mod-Cam will deploy the 280~GHz first light kinetic inductance detector arrays within the first instrument module on FYST for early science observations in 2024. When Prime-Cam is ready for deployment, Mod-Cam will serve as a module testbed for Prime-Cam at Cornell University. In Section \ref{sec:modcam}, we present the design and status of the Mod-Cam cryostat. In Section \ref{sec:280module}, we discuss the design of the 280~GHz instrument module. We detail the 280~GHz detector array design and development status in Section \ref{sec:detectors}, and the readout system design in Section \ref{sec:readout}. We present results from initial tests in Section \ref{sec:initial}, and future plans in Section \ref{sec:future}.

\section{Mod-Cam}\label{sec:modcam}

\begin{figure}
    \centering
    \includegraphics[width=1.0\linewidth]{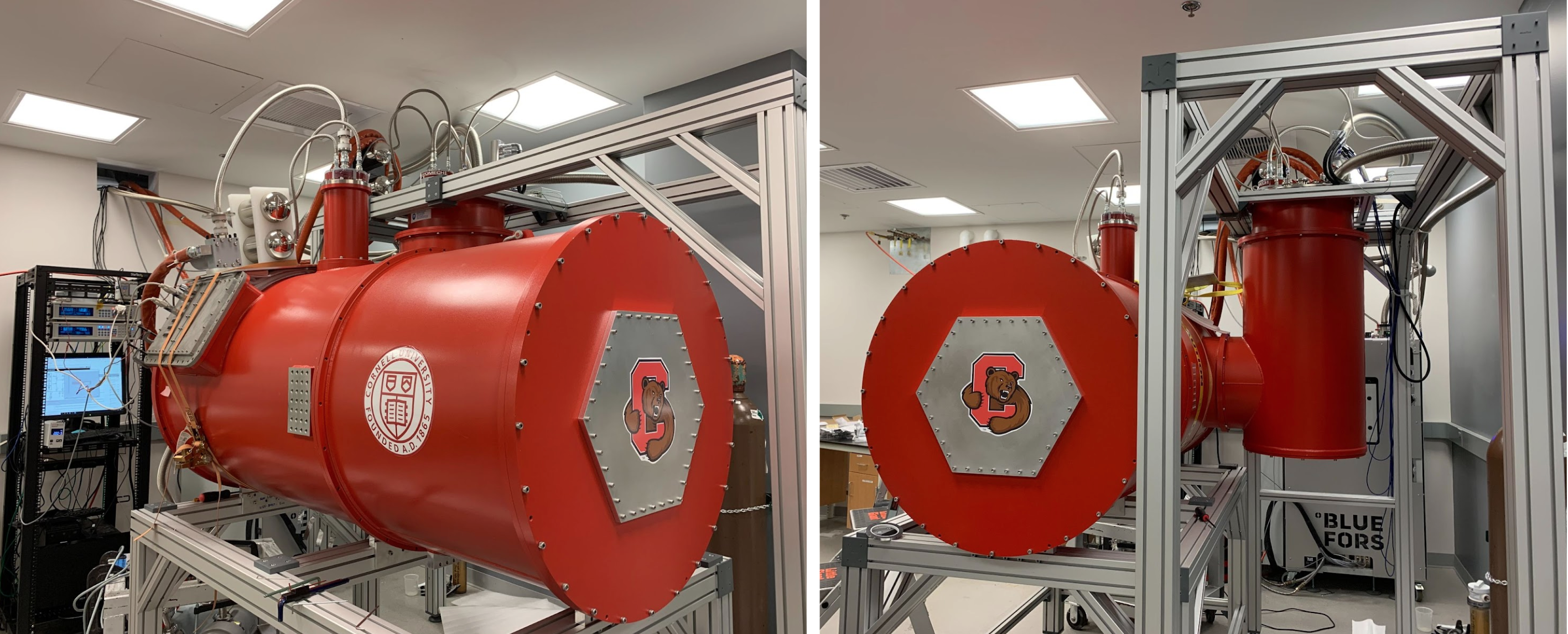}
    \caption{Mod-Cam in testing at Cornell University. In this configuration, a Cryomech PT-420 is installed in addition to the Bluefors LD-400 DR. Mod-Cam sits on a custom Minitec frame. A wheeled frame supporting the main cylinder can be adjustably positioned relative to the DR frame, which can be leveled, raised or lowered using four leveling feet. Left: Thermometry is read out through the readout harness. Right: The side-car DR design is visible. The hexagonal window is blanked off with a piece of 6061-T6 Al for initial cryogenic testing.}
    \label{fig:modcaminlab}
\end{figure}

Mod-Cam is a 0.9 m diameter, 1.8 m long single instrument module cryogenic receiver with 40~K and 4~K stages, currently in testing at Cornell University (Fig.~\ref{fig:modcaminlab}). Mod-Cam will be a first light and commissioning instrument for FYST, and then serve as a testbed for Prime-Cam \cite{EMVSPIE2018}, allowing for optical testing of instruments before they are deployed to the CCAT-prime site for installation in Prime-Cam.

\subsection{Mechanical Design}\label{sec:modcammechanical}

Mod-Cam enables easier swapping of instrument modules as compared to Prime-Cam due to its off-axis dilution refrigerator design and significantly faster turnaround times than the larger cryostat. Each instrument module tested or deployed in Mod-Cam will be optimized for a specific subset of the overall Prime-Cam science goals and be able to hold up to three detector arrays cooled to 100\,mK by a Bluefors LD-400 dilution refrigerator (DR) along with silicon lenses and filter stacks at 1\,K and 4\,K. The instrument modules are mounted to Mod-Cam's 4~K plate. The Mod-Cam cryostat and G10 tabs were fabricated by Precision Cryogenics. 

\begin{figure}
    \centering
    \includegraphics[width=0.9\linewidth]{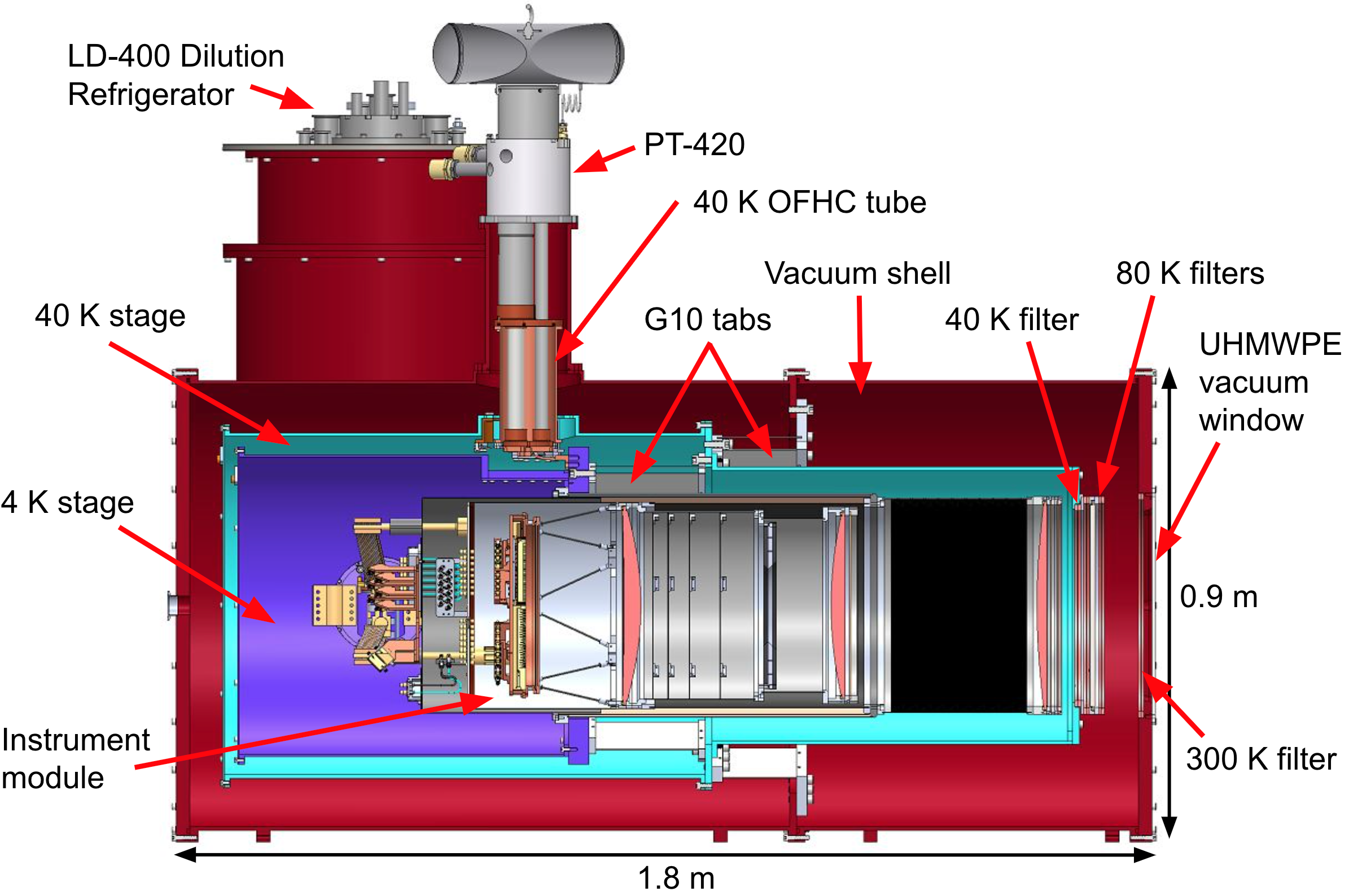}
    \caption{A cross-section view of the Mod-Cam model, showing the interior of the main cylinder of the cryostat which houses the instrument module. Section \ref{sec:modcammechanical} details the design of the 300 K vacuum shell and 40~K and 4~K stages of the cryostat, supported by a series of G-10 tabs and cooled by a Bluefors LD-400 dilution refrigerator and Cryomech PT-420. Light from the telescope passes through an AR-coated UHMWPE window and IR-blocking filters at 300 K and 40~K before entering the instrument module. The filters labeled ``80 K" will be mounted on Prime-Cam's 80 K stage, but are mounted on the 40~K stage in Mod-Cam. The 4K DR adapter thermal connection to the 4~K shell (purple) and DR cold finger connections to the instrument module cold fingers are visible, and shown in detail in Fig.~\ref{fig:DRxsec}. The design of the instrument module is presented in Sec.~\ref{sec:280module} and Fig.~\ref{fig:modulexsec}.}
    \label{fig:modcamhalf}
\end{figure}

The design of the Mod-Cam cryostat portion which houses the instrument module (viewed in cross-section in Fig.~\ref{fig:modcamhalf}) was scaled down from the Prime-Cam cryostat design \cite{EMVSPIE2018}. The off-axis DR shell portion was designed to allow flexible access to both the rear of the cryostat for instrument module installation and removal as well as to the DR. The instrument modules, which can be 41 cm in diameter or smaller, are installed from the back of the cryostat and are cantilevered off of the 4\,K stage. 

\begin{figure}[t!]
    \centering
    \includegraphics[width=0.85\linewidth]{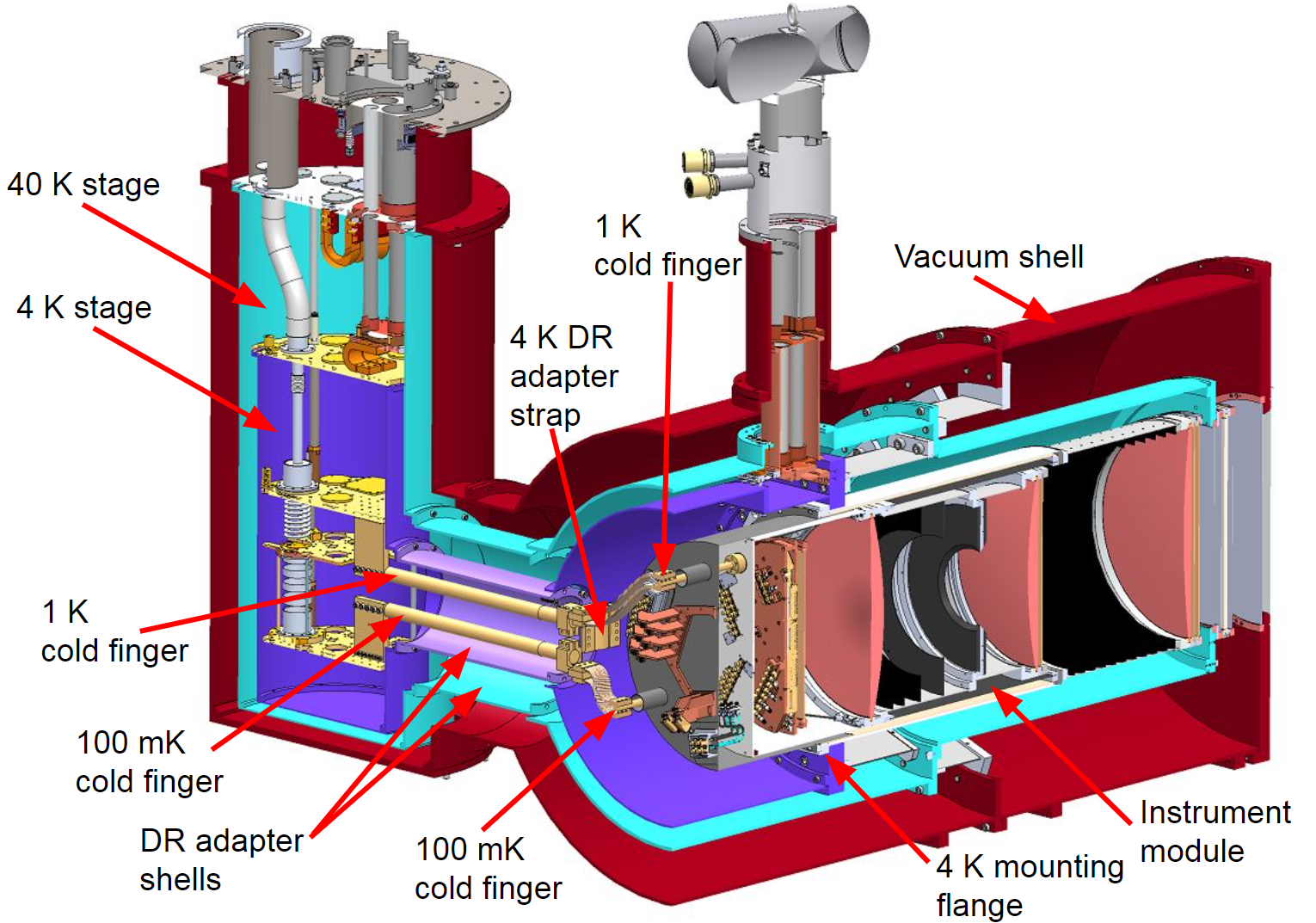}
    \caption{A cutaway view of the Mod-Cam model to reveal the DR to main shell interfaces. 1~K and 100~mK cold fingers attach to the DR plates via OFHC copper blocks and extend into the main cylinder of Mod-Cam to attach to the OFHC 1~K and 100~mK instrument module cold fingers via braided OFHC copper straps from TAI. The 40~K DR adapter cylinder that runs between the 40~K DR shell and the 40~K rear shell is shown in teal. The adapter shell is connected to the 40~K rear shell via two braided OFHC copper straps from TAI (not visible). Similarly, the 4~K DR adapter cylinder that runs between the 4~K DR shell and the 4~K shell is shown in purple, and is connected to the 4~K shell via two braided OFHC copper straps from TAI (one shown).}
    \label{fig:DRxsec}
\end{figure}

Mod-Cam's 6061-T6 Al 300 K vacuum shell consists of 89 cm diameter front and rear shells, front and back plates, a two-piece DR shell, and a DR shell bottom plate (Fig.~\ref{fig:DRxsec}). The size of the optical axis cryostat and the two-piece DR shell were chosen to accommodate one 41 cm diameter optics tube and enough clearance for the Prime-Cam G10 tab design and thermal connections to the cryogenics. The thicknesses of the plates (2.54 cm) and shells (0.64 cm) were motivated by finite-element analysis (FEA) results for the Prime-Cam cryostat \cite{EMVSPIE2018}. The front vacuum plate holds a 44 cm hexagonal ultra-high-molecular-weight polyethylene (UHMWPE) vacuum window, identical to the Simons Observatory Large Aperture Telescope Receiver \cite{Zhu2021} and Prime-Cam designs. The vacuum windows are designed to be thick enough to withstand the atmospheric pressure at the site while being as thin as possible to achieve our desired sensitivity. For site operations, a 0.32 cm (1/8'') thick window, anti-reflection (AR) coated at Cardiff University will be used. A 0.64 cm (1/4'') thick UHMWPE window will be used for laboratory testing to reduce the risk of vacuum window failure \cite{Zhu2021}. Behind the vacuum window lies a double-sided IR-blocking filter fabricated by Cardiff University to reduce loading \cite{Ade2006}. 

Mod-Cam's 40~K stage consists of a 52 cm diameter, 0.64 cm thick front shell, a 65 cm diameter, 0.64 cm thick rear shell, a 1.27 cm thick front filter plate, a 0.32 cm thick back plate, a 1.27 cm thick G-10 mounting ring, a DR shield mounted to the 40~K DR plate, a DR shell to main cylinder adapter, and a DR shell bottom plate \cite{EMVthesis}. All stage components are fabricated of 6063-T5 Al (with the exception of the 6061-T6 Al DR shell bottom plate) for enhanced thermal conductivity and reduction of thermal gradients \cite{Scherer:2018}. The front filter plate holds a filter stack that will be held at 80 K in Prime-Cam, consisting of two double-sided IR-blocking filters fabricated by Cardiff University and one alumina filter. The alumina wedge filters act as IR absorbers, and for off-central tubes in Prime-Cam, as prisms to bend off-axis beams parallel to the long axis of the cryostat such that the instrument modules can all be coaxial with the cryostat shells \cite{dicker:2018}. While Mod-Cam will not see loads necessitating an 80 K plate, the 80 K filters are included at 40~K to test the full optical chain. An additional double-sided IR-blocking filter is mounted to the 40~K plate, which will also be mounted to the 40~K plate in Prime-Cam. The optical elements in Mod-Cam are module-specific and will be swapped out when testing other modules. They are also all compatible with Prime-Cam and will be installed in Prime-Cam along with the tested instrument module. 

Nine 0.24 cm thick, 16 cm by 16 cm G-10 tabs epoxied into 6061-T6 Al feet with Armstrong A-12B PT epoxy by Precision Cryogenics mechanically support the 40~K shells off of the rear 300 K vacuum shell while thermally isolating the 40~K stage (Fig.~\ref{fig:modcamhalf}). Precision Cryogenics sourced the G-10 material from McMaster-Carr\footnote{\url{https://www.mcmaster.com/}}. The design of the G-10 tabs follows from the FEA-motivated designs for the SO LATR \cite{Scherer:2018}. The rear 40~K shell is thermally connected to the 40~K stage of a Cryomech PT-420 via an oxygen-free high thermal conductivity (OFHC) tube and a set of custom compressed OFHC foil straps, as well as to the 40~K DR shell via the 40~K adapter and set of braided OFHC straps from TAI\footnote{\url{https://www.techapps.com/}} (Fig.~\ref{fig:DRxsec}). The rear shell also supports the 40~K stage of the readout harness (Figure \ref{fig:modulereadout}). The 40~K shells and plates are wrapped in 30 custom-cut layers of multi-layer insulation (MLI)\footnote{Beyond Gravity Austria GmbH, Stachegasse 13, 1120 Vienna}, and the G-10 tabs are wrapped in 10 layers of MLI to reduce radiative loading from the room temperature vacuum shell.

Mod-Cam's 4~K stage consists of a 3.8 cm thick plate, a 57 cm diameter, 0.40 cm thick shell, a DR shell mounted to the 4~K plate of the DR, a DR shell to main cylinder adapter, and a DR shell bottom plate, all fabricated from 6061-T6 Al. The 4~K plate is mechanically supported and thermally isolated from the 40~K ring by a series of nine 21.4 cm x 15 cm, 0.24 cm thick G-10 tabs epoxied into 6061-T6 Al feet with Armstrong A-12B PT epoxy by Precision Cryogenics (Fig.~\ref{fig:modcamhalf}). The 4~K plate is the only mechanical mounting point for the instrument modules. The 4~K shell supports the 4~K stage of the readout harness (Figure \ref{fig:modulereadout}). The 4~K plate is thermally connected to the 4~K stage of a Cryomech PT-420 via a set of braided OFHC straps from TAI, and the 4~K DR adapter is also connected to the 4~K shell by two braided OFHC straps from TAI (Fig.~\ref{fig:DRxsec}). The 4~K shells, plates, and G-10 tabs are wrapped in 10 custom-cut layers of MLI to reduce radiative loading from the 40~K stage. 

A mock-up model of the Bluefors LD-400 300, 40, and 4\,K plates and offsets was designed and fabricated for construction of the Mod-Cam cryostat at Precision Cryogenics to ensure the alignment of the DR shells relative to the optical axis Mod-Cam shells. A two-part custom Minitec\footnote{\url{https://www.minitecframing.com}} frame allows for adjustable positioning of the main cylinder and DR shells (Fig.~\ref{fig:modcaminlab}). The side-mounted DR provides cooling to the 40 and 4\,K stages, and an optional Cryomech PT-420 (currently installed) or PT-410  pulse tube can provide cooling power at 40\,K and 4\,K, through custom OFHC braided straps from TAI or straps made in house (Sec.~\ref{sec:cryo}). Thermometry and RF signals are read out through a custom modular harness based on the SO Universal Readout Harness \cite{Moore2022,Rao2020} that is installed on the opposing side to the DR (Sec.~\ref{sec:readout}). The modularity of the harness allows for flexible and upgradable readout options, and its design leaves the rear of Mod-Cam relatively clear for instrument module installation and removal.

\subsection{Cryogenic Design}\label{sec:cryo}

\begin{table}[h!]
\begin{center}
\vspace{2mm}
\begin{tabular}{ |c|c|c|c|c| } 
 \hline
\textbf{Stage} & \textbf{40~K [W]} &\textbf{4~K [W]} & \textbf{1~K [mW]} & \textbf{100~mK [$\mu$W]}\\
\hline
Shell radiation & 9.4 & 0.07 & 0.002 & 0.2\\
Support structure & 3.6 & 0.12 & 0.755 & 16.1\\
Wiring & 5.4 & 0.29 & 0.121 & 3.1\\
Beam radiation & 11.8 & 0.03 & 0.029 & 31.2 \\
\hline
Total heat load & 28.4 & 0.51 & 0.907 & 50.6 \\
\hline
Available cooling power & 110 & 4 & 24 & 400 \\
\hline 
\end{tabular}
\vspace{2mm}
\caption{Loading estimates for each stage of Mod-Cam. The cooling power at 40~K and 4~K is supplied by the PT-420 in the Bluefors LD-400 and the backup Cryomech PT-420, and the cooling power at 1 and 0.1~K is supplied by the DR still and mixing chamber stages respectively. Our estimated cooling power is more than sufficient to meet our estimated needs at all stages for an SO-style instrument module.}
\label{tab:modcamloadingestimates}
\end{center}
\end{table}

The cooling requirements for Mod-Cam are driven by the 100~mK operation of our MKID detectors (Sec.~\ref{sec:detectors}), which are mounted on a stage thermally connected to the 100~mK plate of the LD-400 DR via an instrument module cold finger, flexible TAI straps, DR cold finger and mounting block (Fig.~\ref{fig:DRxsec}). The instrument module 1~K stage is connected to the 1~K DR plate through an analogous chain (Fig.~\ref{fig:DRxsec}). The Mod-Cam shells at 40 and 4~K and the 4~K instrument module stage are cooled by the PT-420 backing up the LD-400 DR as well as by an additional Cryomech PT-420. 

In considering the thermal loads for Prime-Cam and Mod-Cam, we scaled from the SO LATR thermal model \cite{Zhu2021}. This thermal model combines material properties and radiation estimates with custom Python estimates of the optical filter elements. The thermal loading estimates for Mod-Cam are presented in Table \ref{tab:modcamloadingestimates}. The thicknesses and materials of the Mod-Cam shells were determined based on estimated loading and temperature gradients, and the room-temperature mechanical offsets were designed based on anticipated contractions during cooling \cite{EMVthesis}.

\section{280~GHz Instrument Module}\label{sec:280module}

The first instrument module for Mod-Cam and Prime-Cam will be the 280~GHz module, which will contain the first light MKID arrays for the CCAT-prime Project, as described in Section \ref{sec:detectors} and Ref.~\citenum{DuellSPIE}. The instrument module design is based on the optics tube designs for the Simons Observatory LATR \cite{Xu2020,Zhu2021,gallardo:2018,Gudmundsson_2021}, and like all the modules planned for Prime-Cam, is a self-contained assembly of filters, lenses, and detector arrays. The module will be mounted on the 4\,K plate of Mod-Cam (and compatible with Prime-Cam) (Sec.~\ref{sec:modcam}). Each module is approximately 41 cm in diameter and 130 cm long, and mounts through the rear of Mod-Cam (Fig.~\ref{fig:modcamhalf}). 

\subsection{Cold Optics}\label{sec:moduleoptics}

\begin{figure}
    \centering
    \includegraphics[width=1.0\linewidth]{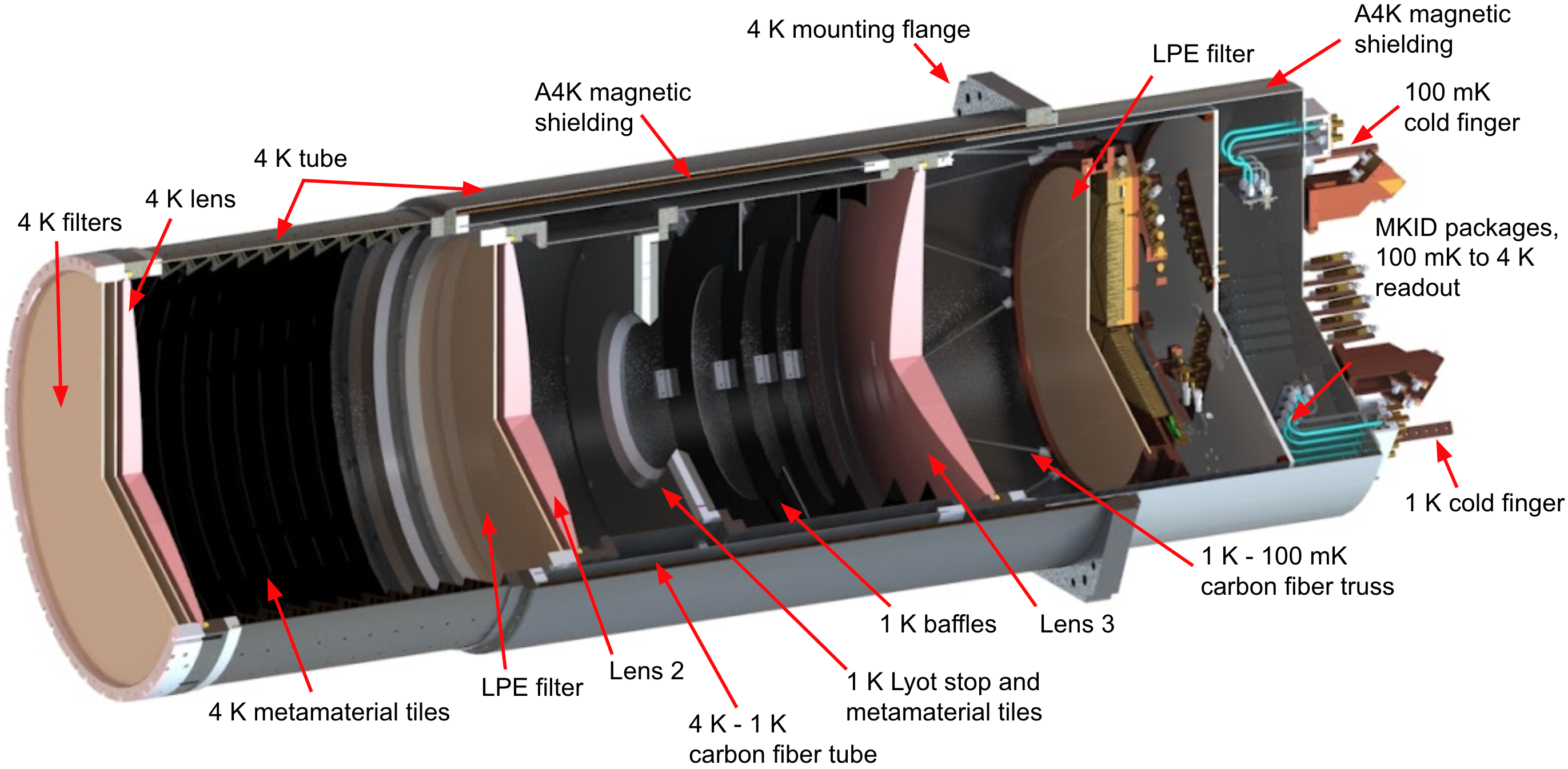}
    \caption{A section view of the 280~GHz instrument module model, showing the 4~K, 1~K and 100~mK stages holding three silicon lenses and LPE filters. The details of the design are presented in Sec.~\ref{sec:280module}. Metamaterial tiles on the upper 4~K tube (Fig.~\ref{fig:baffles}) and 1~K Lyot stop serve as antireflection coatings to absorb stray light, as do the 1~K baffles and shell coated in carbon-loaded epoxy (Fig.~\ref{fig:baffles}). The detector array packages detailed in Sec.~\ref{sec:detectors} are shown, offset from the 1~K stage by a carbon fiber truss. The 100~mK to 4~K readout detailed in Sec.~\ref{sec:readout} is shown in the back along with the 1~K and 100~mK cold fingers which thermally connect to the cold stages of the DR inside Mod-Cam (Fig.~\ref{fig:DRxsec}).}
    \label{fig:modulexsec}
\end{figure}

The design of the 280~GHz module is shown in Fig.~\ref{fig:modulexsec}. Light entering Mod-Cam is filtered through the AR-coated UHMWPE window, 300 K IR-blocking filter, 80 K IR-blocking filters and alumina wedge (located on the 40~K stage in Mod-Cam), and 40~K IR-blocking filter to reduce the loading on the colder stages (as described in Sec.~\ref{sec:modcam}) before entering the instrument module. At 4~K, two low-pass edge (LPE) filters (capacitive mesh filters on polypropylene substrates\cite{Ade2006}) manufactured at Cardiff University are mounted before the first lens of the module.

The cold optics chain continues at 4~K with the first of three metamaterial AR-coated silicon lenses which re-image the telescope focal plane onto the detector arrays. The optical design for the 280~GHz module is adopted from the SO LATR designs \cite{dicker:2018,Zhu2021}. Silicon is the preferred lens material at Prime-Cam's desired wavelengths due to its high resistivity, extremely low loss, high thermal conductivity (ensuring lens temperature uniformity and limiting detector
background loading), and high index of refraction \cite{dicker:2018,Zhu2021,EMVSPIE2018}. Mechanically robust metamaterial antireflection lens coatings applied with a custom CNC machine produce less than 1$\%$ reflection across an octave of bandwidth \cite{Golec2020,Coughlin2018,datta2013}. The second and third lenses are cooled to 1~K. One additional LPE is mounted before the second lens at 1~K and one before the feedhorn arrays at 100~mK (Fig.~\ref{fig:modulexsec}). 

\begin{figure}
    \centering
    \includegraphics[width=1.0\linewidth]{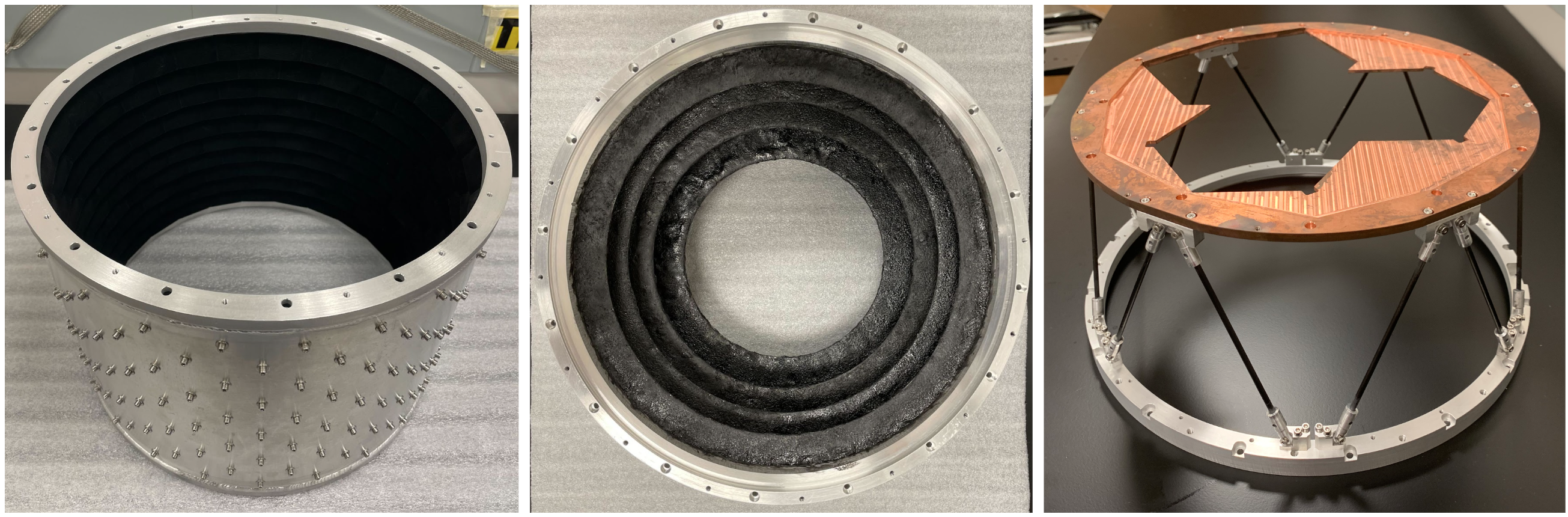}
    \vspace{2mm}
    \caption{Left: The $\sim$240 metamaterial injection-molded plastic tiles\cite{Xu2021,Zhu2021} mounted to the upper 4~K shell of the 280~GHz module to absorb stray light. Center: 1~K baffles installed in lower 1~K tube and coated in carbon-loaded epoxy. Right: 1~K to 100~mK carbon fiber truss.}
    \label{fig:baffles}
\end{figure}

To mitigate stray light inside the module, the injection molded, carbon-loaded plastic metamaterial tile coating design developed for the SO LATR modules was adopted \cite{Xu2021,Gudmundsson_2021,Zhu2021}. Approximately 240 wedge-shaped metamaterial tiles are installed in the upper 4~K section of the instrument module to absorb stray light before the 1~K optics (Fig.~\ref{fig:baffles}, left). The 1~K Lyot stop at the module's pupil is also coated in flat versions of these metamaterial absorbing tiles. After the stop, a series of 1~K ring baffles and the surrounding 1~K shield are coated with Stycast
2850 FT, loaded with coarse and fine carbon powder (Fig.~\ref{fig:baffles}, center). This stage of blackening is less critical for the optical design because of the larger radial beam clearance and position in the module \cite{Zhu2021}. The final lens then focuses the light onto the detector feedhorn arrays (Sec.~\ref{sec:detectors}).

\subsection{Mechanical Design}\label{sec:modulemech}

The 280~GHz instrument module, which contains 4~K, 1~K, and 100~mK stages, is mechanically supported by the 4~K mounting flange on the 4~K plate of Mod-Cam. The Prime-Cam 4~K plate is designed to mount up to seven of the same instrument modules with 4~K flanges. The LPE filters are mounted in 6061-T6 Al clamps axially spring-loaded with Spira\footnote{\url{www.spira-emi.com}} gaskets to reduce the likelihood of delamination via thermal contraction\cite{Zhu2021}. Each lens clamp also supports the lens with axial and radial Spira gaskets. The lens clamps were modified from the SO LATR tube design to include a finger hole and block for ease of assembly when installing the silicon lenses. 

The 4~K components of the module include the welded 6061-T6 Al upper tube which is covered with 4~K metamaterial wedge-shaped tiles (Fig.~\ref{fig:baffles}, left) and the welded 6061-T6 Al lower tube which includes the 4~K mounting flange and mounting point for the 4~K magnetic shielding. The A4K\footnote{Amuneal 4K material, \url{www.amuneal.com/}} magnetic shielding for the module was developed from the SO LATR design to accommodate the 280~GHz MKID readout design (Sec.~\ref{sec:readout}), and extends along the interior of the lower 4~K tube past the second lens, as well as around the outside of the rear of the module. Laboratory testing of superconducting detectors and readout components like the TESes and SQUIDs used for SO motivated the design of this shield\cite{VavagiakisASC2021}. MKIDs are anticipated to be less sensitive to magnetic fields than TESes or SQUIDs, but initial magnetic sensitivity testing has illustrated the importance of shielding MKID arrays\cite{Choi2021}.

The 1~K stage of the instrument module is mechanically supported and thermally isolated from the 4~K shells by a 38.54 cm diameter, 0.3 cm thick carbon fiber tube from DragonPlate/Allred and Associates\footnote{www.DragonPlate.com}, epoxied into 6061-T6 Al rings in an alignment jig with Scotchweld 2216 cured at room temperature. The welded 6061-T6 Al upper 1~K tube supports the first LPE filter, second lens, and 1~K Lyot stop, which is coated with flat plastic metamaterial antireflection coating tiles (Sec.~\ref{sec:moduleoptics}). The welded 6061-T6 Al lower tube supports a series of blackened baffles (Fig.~\ref{fig:baffles}, center) and the third lens. 

A carbon fiber truss supports the 100~mK stage off of the rear of the 1~K stage (Fig.~\ref{fig:modulexsec},~\ref{fig:baffles} right). This truss will be composed of 4 mm outer diameter, 3 mm inner diameter carbon fiber rods made with a pultrusion process from vDijk Pultrusion Products\footnote{vDijk Pultrusion Products, Aphroditestraat 24, NL-5047 TW TILBURG, The Netherlands}. The design of the truss is based on the SO LATR tube truss, but is modified to minimize off-axis loading of the carbon fiber tubes by incorporating updated strut end cap designs from the SO Small Aperture Telescopes\cite{Crowley_2022}. The carbon fiber rods are epoxied into 6061-T6 Al feet with Scotch-Weld DP2216 following the procedure in Crowley at al. 2022\cite{Crowley_2022}. FEA performed on this new design predicts that this truss will support at least five times the expected operating load, and load testing of individual epoxied pultruded carbon fiber struts sourced from Aopin -confirm the FEA predictions, with results showing that each strut can support at least $10\times$ its expected load. SolidWorks modal analysis predicts that the lowest vibrational mode for this truss design will be above 200 Hz. The 100~mK stage supports the feedhorn and detector array packages described in Sec.~\ref{sec:detectors}. The 100~mK stage is surrounded by the 1~K radiation shield. The 1~K radiation shield and 4~K magnetic shielding support the module readout components described in Sec.~\ref{sec:readout}.

The thermometry plan for Mod-Cam involves 18 temperature sensors at important thermal interfaces, locations on plates and shells to probe potential gradients, and within the instrument modules at each temperature stage \cite{EMVthesis}. The number of sensors planned at each temperature stage is presented in Table \ref{tab:thermotable}. Cernox\footnote{\url{shop.lakeshore.com/temperature-products/temperature-sensors/cernox.html}} 1080 thin film resistance cryogenic temperature sensors are selected for the 40\,K stage, Cernox 1050 for the 4\,K stage, Cernox 1030 for the 1\,K stage, and Ruthenium oxide sensors (ROXs) for the 100\,mK stage. LEMO connectors are used for four-lead sensor measurement and read out using Lakeshore resistance bridges. Custom cables from Universal Cryo, Inc.\footnote{\url{www.ucryo.com}} have been ordered for optical testing and deployment. 

\begin{table}
\begin{center}
\begin{tabular}{ |c|c|c|c|c|c| } 
 \hline
Stages & 40\,K & 4\,K & 1\,K & 100\,mK & Total\\
\hline
Thermometers & 6 & 7 & 2 & 3 & 18\\
\hline
\end{tabular}
\vspace{2mm}
\caption{Planned number of thermometers for each stage in Mod-Cam. Six 40\,K thermometers will measure temperature at the 40\,K PT stage, 40\,K DR adapter, and across the 40\,K shells and plates. Seven 4\,K thermometers will measure temperature at the 4\,K PT stage, 4\,K DR adapter, 4\,K instrument module components, and across the 4\,K shells and plates. Two 1\,K and three 100\,mK thermometers will measure temperatures in the instrument module. Cernox sensors will be used at 1\,K and above, while ROXs will be used at 100\,mK.}
\label{tab:thermotable}
\end{center}
\end{table}

\section{280~GHz Detector Arrays}\label{sec:detectors}

In order to meet the desired instrument sensitivity and required detector densities, all of the currently planned instrument modules for Prime-Cam will use microwave kinetic inductance detectors (MKIDs). Signals are measured by coupling incident photons to a superconducting inductive element of an LC resonator and measuring the shift in kinetic inductance caused by Cooper-pair breaking. By combining the absorbing and inductive elements, MKIDs are naturally frequency multiplexed, and have greatly reduced readout complexity in comparison to that required for similarly sensitive transition edge sensors. 

\begin{figure}[ht!]
\centering
    \includegraphics[width=0.8\linewidth]{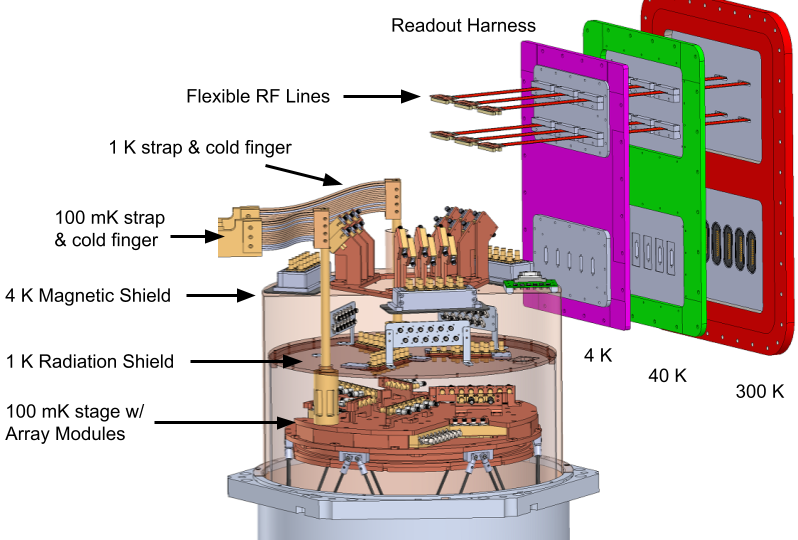}
    \caption{Partial cutaway view of the cryogenic readout for Mod-Cam, including the readout harness, 280~GHz instrument module, and cold straps by TAI. For simplicity, coaxial cables and some transitional elements are not shown. Flexible RF stripline carries both input and output signals for all 18 networks between 300 K and 4~K, where they are transitioned to a combination of flexible and semi-rigid coaxial cables down to the three arrays at 100~mK. 18 low noise amplifiers are mounted and heatsunk on the rear face of the 4~K magnetic shield.}
    \label{fig:modulereadout}
\end{figure}

Within the first-light 280~GHz module, there will be more than 10,000 feedhorn-coupled, polarization-sensitive MKIDs divided between three tiled array packages with two unique designs. These include a first array of TiN detectors coupled to aluminum feedhorns and a further two arrays of Al detectors coupled to silicon feedhorns. The first array has been described previously in Ref.~\citenum{DuellSPIE}. It contains 3,456 total detectors (1,728 pixels), of which 3,450 are optically coupled. The array was fabricated by the Quantum Sensors Group at the National Institute for Standards and Technology, drawing on previous work done for BLAST-TNG\cite{dober_optical_2016, Galitzki2016} and TolTEC\cite{austermann_millimeter-wave_2018, austermann_large_2018}, while the aluminum feedhorns were machined at ASU. Two additional arrays using updated Al MKID designs are currently being fabricated and tested, along with Si-platelet feedhorn arrays, also by the Quantum Sensors Group at NIST\cite{AustermannSPIE22}. Each of these arrays will have 3,448 total detectors (1,724 pixels), of which 3,418 are optically coupled. This change from TiN to Al detectors was driven by dark testing results demonstrating reduced low frequency spectral noise\cite{AustermannSPIE22}. The array packaging differs slightly between both designs to accommodate the different alignment and heat-sinking requirements of the two feedhorn types, though overall pixel spacing and placement is the same. One advantage of this mixed design is the opportunity to sample the same sky using two sets of detectors with unique noise properties, though this does add complexity to data analysis at these frequencies.

\section{Readout System}\label{sec:readout}

The 280~GHz module cryogenic readout design is the first for the Prime-Cam module development effort, and enables the readout of more than 10,000 KIDs ($\sim$3,500 per array) across 18 networks. Each individual 280~GHz array is split into 6 networks with either 576 or 572 resonators placed at frequencies roughly between 500 MHz and 1 GHz to be measured over a single RF feedline. The readout of a fully populated 280~GHz module with three arrays will require 18 pairs of RF feedlines and accompanying low noise amplifiers (LNAs). The room temperature microwave frequency multiplexed readout system for Mod-Cam and Prime-Cam is currently in development, and is designed to run on the Xilinx ZCU111 Radio Frequency System on a Chip  (RFSoC)\cite{Sinclair2022}. In both Mod-Cam and Prime-Cam, the cryogenic readout is broken up between a shared readout harness spanning 300 K to 4~K, individual instrument modules with stages from 4~K to 100~mK, and an isothermal 4~K transition to connect the two. While the instrument modules are fully shared between both receivers, the readout harness and isothermal transitions are modified in Mod-Cam to accommodate its unique layout and purpose as a flexible testbed.

The readout design for the module is shown in Fig.~\ref{fig:modulereadout}. The readout from the 100~mK arrays to 4~K relies on a combination of semi-rigid and hand-formable coaxial cables, with the hand-formable cables being used at isothermal stretches to reduce complexity during installation. Attenuation is included at each stage on the input side to reach the desired tone power and noise temperature, and low-loss superconducting cables carry the output signal across all temperature stages between the array and low noise amplification at 4~K. Coaxial cables running from the focal plane arrays are heat sunk at 1~K on the 1~K radiation shield, as well as at 4~K on the magnetic shield where all LNAs are located. After being routed through the magnetic shield, coaxial cables are surrounded by slotted A4K covers to complete the magnetic shielding. PCBs for breakout of LNA bias lines are also located at 4~K.

From 4~K to 300 K, an RF stripline design based on those used for ALPACA\cite{VishwasSPIE22} runs roughly 18 inches of flexible RF stripline through a readout harness with mechanical designs based on the Universal Readout Harness for the Simons Observatory\cite{Moore2022,Rao2020} (Fig.~\ref{fig:modulereadout}). Each flexible board holds 6 RF feedlines with custom SMP connectors on both ends. These SMP connectors then mate to a transition board that switches all lines to SMA connectors. The readout harness design shown, which is specific to Mod-Cam, sacrifices some efficiency in stripline density to allow for greater modularity when testing modules with alternative readout requirements or additional DC line requirements. Not shown in detail is coaxial cable routing required for transitioning between the readout harness and instrument module. This will require the most substantial modification between Mod-Cam and Prime-Cam, and, as such, is still being finalized.

\section{Initial Tests}\label{sec:initial}

\begin{figure}[t!]
\centering
    \includegraphics[width=0.9\linewidth]{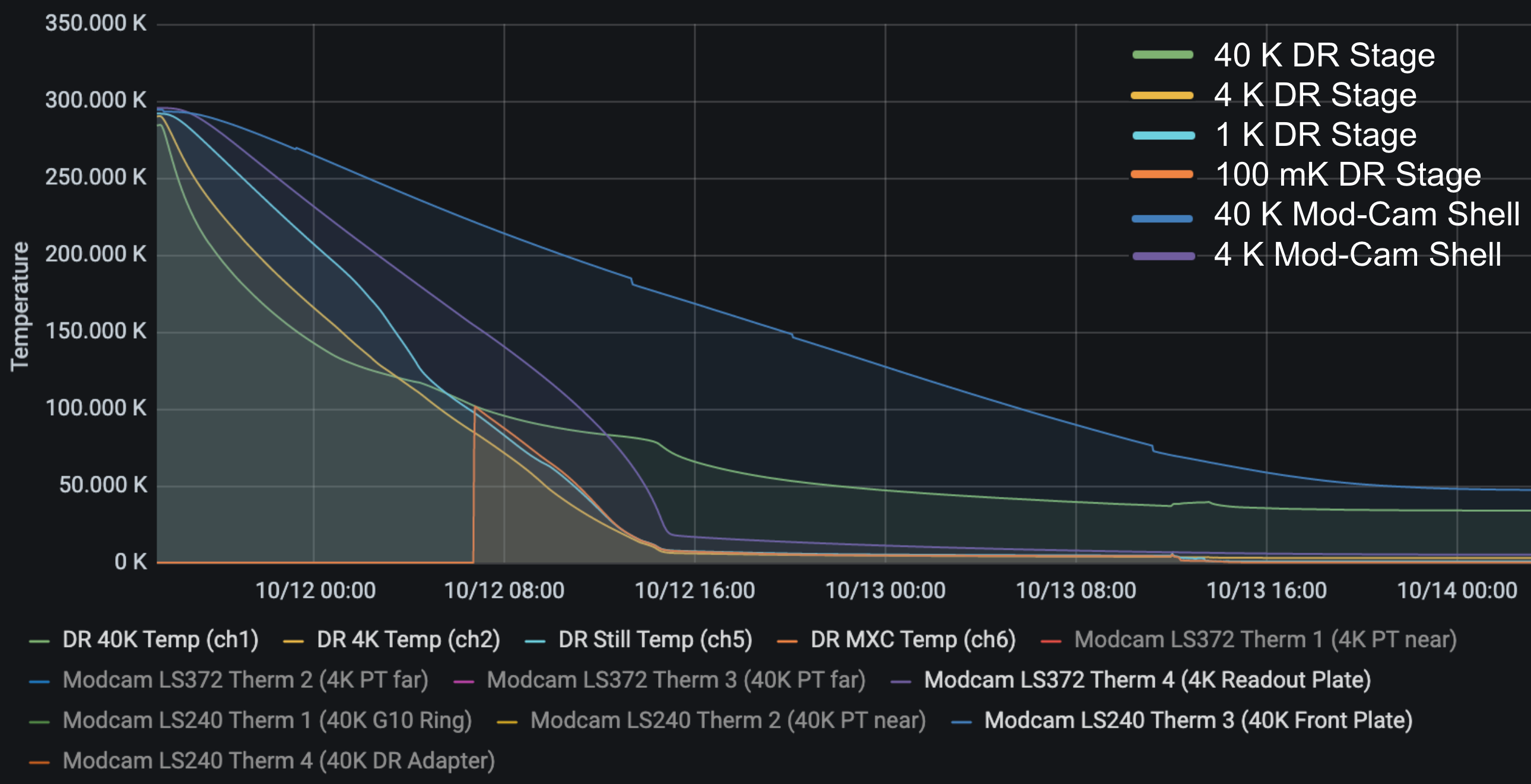}
    \caption{A cooldown plot from Grafana for a dark test of the Mod-Cam receiver in which no instrument module is installed and the window plate, 40~K and 4~K plates are blanked off. A subset of the thermometers installed are plotted for clarity, including thermometers on the 40~K, 4~K, 1~K and 100~mK DR stages and the 40~K and 4~K thermometers on the Mod-Cam shells farthest from cryogenic thermal connections. Base temperature for the DR cold stages at 1~K and 100~mK is reached in roughly 8 hours after the DR is turned on, and the warmer stages reach base in approximately 2.5 days after the two PT-420s are turned on.}
    \label{fig:cooldown}
\end{figure}

Dark testing of the Mod-Cam cryogenic receiver without an instrument module installed has been performed. An initial DR-only cooldown led to a slower cooldown rate than anticipated, so we added a PT-410, and later a PT-420 for comparison. The performance of the DR was tested in a configuration in which a PT-410 was coupled to the 40 and 4\,K main shells of Mod-Cam, and the 40 and 4\,K DR adapters were uncoupled from the main shells such that the DR was not cooling the main shells. In this configuration, the DR mixing chamber achieved over 400 $\mu$W of cooling power at 100\,mK, which is more than sufficient for one LATR-style instrument module \cite{Xu2020}.

Live monitoring of Mod-Cam's pressures and temperatures during cooldowns is achieved using Grafana\footnote{\url{https://grafana.com}}, an open source web application for visualization of time series data, via the Observatory Control System (OCS) used by the Simons Observatory\cite{Koopman2020}. A plot of a test cooldown from Grafana is shown in Fig.~\ref{fig:cooldown} for a dark test configuration in which no instrument is installed, the window is blanked off (Fig.~\ref{fig:modcaminlab}), and the 40 and 4~K Mod-Cam plates are covered. In this configuration, base temperature is reached approximately 3.5 days after the two PT-420s are turned on. The DR cold stages at 1~K and 100~mK reach base temperature in roughly 8 hours after the DR is turned on. 

Lab characterization of the aluminum feedhorn-coupled 280~GHz TiN polarimetric MKID array is ongoing at Cornell\cite{Choi2021}. An LED mapper array PCB has been developed to map MKID resonator frequencies to the physical array positions, in order to lithographically trim the interdigitated capacitors and remove frequency collisions \cite{Liu2017}.

\section{Status and Future Plans}\label{sec:future}

The 280~GHz module is currently under construction at Cornell University and will shortly be dark tested in Mod-Cam for the first time to test base temperatures and thermal gradients present in the module without optical loading, as well as robustness of the module's epoxied components and coatings after thermal cycling. Modal analysis of the Mod-Cam cryostat is ongoing, and vibration tests with a Buttkicker Mini Concert haptic transducer\footnote{\url{https://thebuttkicker.com/buttkicker-mini-concert}} are planned following the method outlined in Ref.~\citenum{Zhu2021} to test the heating of the cold stages from pickup of anticipated low-frequency modes that will be present in the vibrational environment on FYST.

Characterization of the first Al feedhorn-coupled 280~GHz TiN MKID array is ongoing, and two more 280~GHz MKID arrays using aluminum inductors and silicon feedhorns are being fabricated and tested at NIST. Full characterization of the 280~GHz arrays will be presented in a future publication. Optical testing of the first MKID arrays in the 280~GHz module will follow dark testing, with the goal of early science observations on FYST in 2024. 

\section{Conclusion}

The Mod-Cam cryogenic receiver will be a first light and commissioning instrument for the Fred Young Submillimeter Telescope (FYST), deploying the first light MKID arrays at 280~GHz for the CCAT-prime Project. Mod-Cam will also serve as the instrument module testbed for Prime-Cam, a first-generation science instrument for FYST that will perform unprecedented 280--850 GHz broadband and spectroscopic measurements with KIDs. The 0.9 m diameter, 1.8 m long Mod-Cam receiver with 40~K and 4~K stages is currently in testing at Cornell University, and the 41 cm diameter 280~GHz instrument module with cold stages at 40~K, 4~K, 1~K, and 100~mK, is currently under construction for initial cryogenic testing. The first 280~GHz MKID array is currently in testing in the lab, and the readout of more than 10,000 KIDs across 18 networks has been designed. Mod-Cam will be installed on FYST for early science observations in 2024.

\section{Acknowledgements}
The CCAT-prime Project, FYST and Prime-Cam instrument have been supported by generous contributions from the Fred M. Young, Jr. Charitable Trust, Cornell University, and the Canada Foundation for Innovation and the Provinces of Ontario, Alberta, and British Columbia. The construction of the FYST telescope was supported by the Gro{\ss}ger{\"a}te-Programm of the German Science Foundation (Deutsche Forschungsgemeinschaft, DFG) under grant INST 216/733-1 FUGG, as well as funding from Universit{\"a}t zu K{\"o}ln, Universit{\"a}t Bonn and the Max Planck Institut f{\"u}r Astrophysik, Garching. The construction of EoR-Spec is supported by NSF grant AST-2009767. The construction of the 350 GHz instrument module for Prime-Cam is supported by NSF grant AST-2117631. MDN acknowledges support from NSF grant AST-2117631. SKC acknowledges support from NSF award AST2001866. ZBH acknowledges support from a NASA Space Technology Graduate Research Opportunities Award. ZX is supported by the Gordon and Betty Moore Foundation through grant GBMF5215 to the Massachusetts Institute of Technology.

\bibliography{report} 

\begin{thebibliography}{10}

\bibitem{CCATscience}
Aravena, M., Austermann, J.~E., Basu, K., Battaglia, N., Beringue, B.,
  Bertoldi, F., Bigiel, F., Bond, J.~R., Breysse, P.~C., Broughton, C., Bustos,
  R., Chapman, S.~C., Charmetant, M., Choi, S.~K., Chung, D.~T., Clark, S.~E.,
  Cothard, N.~F., Crites, A.~T., Dev, A., Douglas, K., Duell, C.~J., Ebina, H.,
  Erler, J., Fich, M., Fissel, L.~M., Foreman, S., Gao, J., García, P.,
  Giovanelli, R., Haynes, M.~P., Hensley, B., Herter, T., Higgins, R., Huber,
  Z., Hubmayr, J., Johnstone, D., Karoumpis, C., Keating, L.~C., Komatsu, E.,
  Li, Y., Magnelli, B., Matthews, B.~C., Meerburg, P.~D., Meyers, J.,
  Muralidhara, V., Murray, N.~W., Niemack, M.~D., Nikola, T., Okada, Y.,
  Riechers, D.~A., Rosolowsky, E., Roy, A., Sadavoy, S.~I., Schaaf, R.,
  Schilke, P., Scott, D., Simon, R., Sinclair, A.~K., Sivakoff, G.~R., Stacey,
  G.~J., Stutz, A.~M., Stutzki, J., Tahani, M., Thanjavur, K., Timmermann,
  R.~A., Ullom, J.~N., van Engelen, A., Vavagiakis, E.~M., Vissers, M.~R.,
  Wheeler, J.~D., White, S. D.~M., Zhu, Y., and Zou, B., ``{CCAT}-prime
  {C}ollaboration: {S}cience {G}oals and {F}orecasts with {P}rime-{C}am on the
  {F}red {Y}oung {S}ubmillimeter {T}elescope,''  arXiv:2107.10364 (2021).

\bibitem{Parshley2022}
Parshley, S. and et~al., ``{CCAT}-prime: {T}he {F}red {Y}oung {S}ubmillimeter
  {T}elescope {F}inal {D}esign and {F}abrication,'' {\em Paper No. 12182-53}
  (2022).

\bibitem{EMVSPIE2018}
{Vavagiakis}, E.~M., {Ahmed}, Z., {Ali}, A., {Basu}, K., {Battaglia}, N.,
  {Bertoldi}, F., {Bond}, R., {Bustos}, R., {Chapman}, S.~C., {Chung}, D.,
  {Coppi}, G., {Cothard}, N.~F., {Dicker}, S., {Duell}, C.~J., {Duff}, S.~M.,
  {Erler}, J., {Fich}, M., {Galitzki}, N., {Gallardo}, P.~A., {Henderson},
  S.~W., {Herter}, T.~L., {Hilton}, G., {Hubmayr}, J., {Irwin}, K.~D.,
  {Koopman}, B.~J., {McMahon}, J., {Murray}, N., {Niemack}, M.~D., {Nikola},
  T., {Nolta}, M., {Orlowski-Scherer}, J., {Parshley}, S.~C., {Riechers},
  D.~A., {Rossi}, K., {Scott}, D., {Sierra}, C., {Silva-Feaver}, M., {Simon},
  S.~M., {Stacey}, G.~J., {Stevens}, J.~R., {Ullom}, J.~N., {Vissers}, M.~R.,
  {Walker}, S., {Wollack}, E.~J., {Xu}, Z., and {Zhu}, N., ``{Prime-Cam: a
  first-light instrument for the CCAT-prime telescope},'' in [{\em Millimeter,
  Submillimeter, and Far-Infrared Detectors and Instrumentation for Astronomy
  IX}{\nolinebreak\hspace{0.1em}]},  {Zmuidzinas}, J. and {Gao}, J.-R., eds.,
  {\em Society of Photo-Optical Instrumentation Engineers (SPIE) Conference
  Series} {\bf 10708},  107081U (July 2018).

\bibitem{ChoiPrimeCam}
{Choi}, S.~K., {Austermann}, J., {Basu}, K., {Battaglia}, N., {Bertoldi}, F.,
  {Chung}, D.~T., {Cothard}, N.~F., {Duff}, S., {Duell}, C.~J., {Gallardo},
  P.~A., {Gao}, J., {Herter}, T., {Hubmayr}, J., {Niemack}, M.~D., {Nikola},
  T., {Riechers}, D., {Rossi}, K., {Stacey}, G.~J., {Stevens}, J.~R.,
  {Vavagiakis}, E.~M., {Vissers}, M., and {Walker}, S., ``{Sensitivity of the
  Prime-Cam Instrument on the CCAT-Prime Telescope},'' {\em Journal of Low
  Temperature Physics}~{\bf 199},  1089--1097 (Mar. 2020).

\bibitem{SO:2019}
{Simons Observatory Collaboration}, ``{{T}he {S}imons {O}bservatory: {S}cience
  {G}oals and {F}orecasts},'' {\em Journal of Cosmology and Astroparticle
  Physics}~{\bf 2019},  056 (Feb. 2019).

\bibitem{Xu2021}
Xu, Z., Chesmore, G.~E., Adachi, S., Ali, A.~M., Bazarko, A., Coppi, G.,
  Devlin, M., Devlin, T., Dicker, S.~R., Gallardo, P.~A., Golec, J.~E.,
  Gudmundsson, J.~E., Harrington, K., Hattori, M., Kofman, A., Kiuchi, K.,
  Kusaka, A., Limon, M., Matsuda, F., McMahon, J., Nati, F., Niemack, M.~D.,
  Suzuki, A., Teply, G.~P., Thornton, R.~J., Wollack, E.~J., Zannoni, M., and
  Zhu, N., ``The {S}imons {O}bservatory: metamaterial microwave absorber and
  its cryogenic applications,'' {\em Applied Optics}~{\bf 60},  864 ({J}an
  2021).

\bibitem{Zhu2021}
Zhu, N., Bhandarkar, T., Coppi, G., Kofman, A.~M., Orlowski-Scherer, J.~L., Xu,
  Z., Adachi, S., Ade, P., Aiola, S., Austermann, J., Bazarko, A.~O., Beall,
  J.~A., Bhimani, S., Bond, J.~R., Chesmore, G.~E., Choi, S.~K., Connors, J.,
  Cothard, N.~F., Devlin, M., Dicker, S., Dober, B., Duell, C.~J., Duff, S.~M.,
  Dünner, R., Fabbian, G., Galitzki, N., Gallardo, P.~A., Golec, J.~E.,
  Haridas, S.~K., Harrington, K., Healy, E., Ho, S.-P.~P., Huber, Z.~B.,
  Hubmayr, J., Iuliano, J., Johnson, B.~R., Keating, B., Kiuchi, K., Koopman,
  B.~J., Lashner, J., Lee, A.~T., Li, Y., Limon, M., Link, M., Lucas, T.~J.,
  McCarrick, H., Moore, J., Nati, F., Newburgh, L.~B., Niemack, M.~D.,
  Pierpaoli, E., Randall, M.~J., Sarmiento, K.~P., Saunders, L.~J., Seibert,
  J., Sierra, C., Sonka, R., Spisak, J., Sutariya, S., Tajima, O., Teply,
  G.~P., Thornton, R.~J., Tsan, T., Tucker, C., Ullom, J., Vavagiakis, E.~M.,
  Vissers, M.~R., Walker, S., Westbrook, B., Wollack, E.~J., and Zannoni, M.,
  ``The {S}imons {O}bservatory {L}arge {A}perture {T}elescope {R}eceiver,''
  {\em The Astrophysical Journal Supplement Series}~{\bf 256},  23 (sep 2021).

\bibitem{DuellSPIE}
{Duell}, C.~J., {Vavagiakis}, E.~M., {Austermann}, J., {Chapman}, S.~C.,
  {Choi}, S.~K., {Cothard}, N.~F., {Dober}, B., {Gallardo}, P., {Gao}, J.,
  {Groppi}, C., {Herter}, T.~L., {Stacey}, G.~J., {Huber}, Z., {Hubmayr}, J.,
  {Johnstone}, D., {Li}, Y., {Mauskopf}, P., {McMahon}, J., {Niemack}, M.~D.,
  {Nikola}, T., {Rossi}, K., {Simon}, S., {Sinclair}, A.~K., {Vissers}, M.,
  {Wheeler}, J., and {Zou}, B., ``{CCAT-prime: Designs and status of the first
  light 280 GHz MKID array and Mod-Cam receiver},'' in [{\em Society of
  Photo-Optical Instrumentation Engineers (SPIE) Conference
  Series}{\nolinebreak\hspace{0.1em}]},  {\em Society of Photo-Optical
  Instrumentation Engineers (SPIE) Conference Series} {\bf 11453},  114531F
  (Dec. 2020).

\bibitem{Choi2021}
Choi, S.~K., Duell, C.~J., Austermann, J., Cothard, N.~F., Gao, J., Freundt,
  R.~G., Groppi, C., Herter, T., Hubmayr, J., Huber, Z.~B., Keller, B., Li, Y.,
  Mauskopf, P., Niemack, M.~D., Nikola, T., Rossi, K., Sinclair, A., Stacey,
  G.~J., Vavagiakis, E.~M., Vissers, M., Tucker, C., Weeks, E., and Wheeler,
  J., ``{CCAT-prime}: {C}haracterization of the first 280 {GH}z {MKID} array
  for {P}rime-{C}am,'' (2021).

\bibitem{Ade2006}
Ade, P., Pisano, G., Tucker, C., and Weaver, S., ``A review of metal mesh
  filters - art. no. 62750u,'' {\em Proceedings of SPIE - The International
  Society for Optical Engineering}~{\bf 6275} (07 2006).

\bibitem{EMVthesis}
Vavagiakis, E.~M., {\em {Measuring the Sunyaev-Zel'dovich Effects with Current
  and Future Observatories}}, PhD thesis, Cornell University (2021).

\bibitem{Scherer:2018}
Orlowski-Scherer, J.~L., Zhu, N., Zhu, X., Arnold, K.~S., Simon, S.~M.,
  Galitzki, N., Dikcer, S., Limon, M., Devlin, M.~J., Niemack, M.~D., Puglisi,
  G., Coppi, G., Vavagiakis, E.~M., Silva-Feaver, M., Keating, B., Ali, A.,
  Piccirillo, L., Lee, A.~T., Gallardo, P.~A., Salatino, M., Ashton, P.~C.,
  McMahon, J., Lungu, M., May, A.~J., and Thornton, R., ``Simons {O}bservatory
  {L}arge {A}perture {R}eceiver {S}imulation {O}verview,'' {\em Millimeter,
  Submillimeter, and Far-Infrared Detectors and Instrumentation for Astronomy
  {IX}}  ({J}ul 2018).

\bibitem{dicker:2018}
Dicker, S.~R., Gallardo, P.~A., Gudmundsson, J.~E., Mauskopf, P.~D., Ali, A.,
  Ashton, P.~C., Coppi, G., Devlin, M.~J., Galitzki, N., Ho, S.~P., Hill,
  C.~A., Hubmayr, J., Keating, B., Lee, A.~T., Limon, M., Matsuda, F., McMahon,
  J., Niemack, M.~D., Orlowski-Scherer, J.~L., Piccirillo, L., Salatino, M.,
  Simon, S.~M., Staggs, S.~T., Thornton, R., Ullom, J.~N., Vavagiakis, E.~M.,
  Wollack, E.~J., Xu, Z., and Zhu, N., ``Cold optical design for the {L}arge
  {A}perture {S}imons {O}bservatory {T}elescope,'' (2018).

\bibitem{Moore2022}
Moore, J.~E., Bhandarkar, T., DiGia, B., Duell, C., Galitzki, N., Mathewson,
  J., Orlowski-Scherer, J., Silva-Feaver, M., Wang, Y., Wheeler, C., Xu, Z.,
  and Mauskopf, P., ``Development and performance of {U}niversal {R}eadout
  {H}arnesses for the {S}imons {O}bservatory,'' (2022).

\bibitem{Rao2020}
{Sathyanarayana Rao}, M., {Silva-Feaver}, M., et~al., ``{Simons Observatory
  Microwave SQUID Multiplexing Readout: Cryogenic RF Amplifier and Coaxial
  Chain Design},'' {\em Journal of Low Temperature Physics}~{\bf 199},
  807--816 (Mar. 2020).

\bibitem{Xu2020}
{Xu}, Z., {Bhandarkar}, T., {Coppi}, G., {Kofman}, A., {Orlowski-Scherer},
  J.~L., {Zhu}, N., {Ali}, A.~M., {Arnold}, K., {Austermann}, J.~E., {Choi},
  S.~K., {Connors}, J., {Cothard}, N.~F., {Devlin}, M., {Dicker}, S., {Dober},
  B., {Duff}, S.~M., {Fabbian}, G., {Galitzki}, N., {Haridas}, S.,
  {Harrington}, K., {Healy}, E., {Ho}, S.-P.~P., {Hubmayr}, J., {Iuliano}, J.,
  {Lashner}, J., {Li}, Y., {Limon}, M., {Koopman}, B.~J., {McCarrick}, H.,
  {Moore}, J., {Nati}, F., {Niemack}, M.~D., {Reichardt}, C.~L., {Sarmiento},
  K., {Seibert}, J., {Silva-Feaver}, M., {Sonka}, R.~F., {Staggs}, S.,
  {Thornton}, R.~J., {Vavagiakis}, E.~M., {Vissers}, M.~R., {Walker}, S.,
  {Wang}, Y., {Wollack}, E.~J., and {Zheng}, K., ``{The Simons Observatory: the
  Large Aperture Telescope Receiver (LATR) integration and validation
  results},'' in [{\em Society of Photo-Optical Instrumentation Engineers
  (SPIE) Conference Series}{\nolinebreak\hspace{0.1em}]},  {\em Society of
  Photo-Optical Instrumentation Engineers (SPIE) Conference Series} {\bf
  11453},  1145315 (Dec. 2020).

\bibitem{gallardo:2018}
Gallardo, P.~A. et~al., ``{Studies of Systematic Uncertainties for Simons
  Observatory: Optical Effects and Sensitivity Considerations},'' {\em Proc.
  SPIE} (10708-133) (2018).

\bibitem{Gudmundsson_2021}
Gudmundsson, J.~E., Gallardo, P.~A., Puddu, R., Dicker, S.~R., Adler, A.~E.,
  Ali, A.~M., Bazarko, A., Chesmore, G.~E., Coppi, G., Cothard, N.~F.,
  Dachlythra, N., Devlin, M., Dünner, R., Fabbian, G., Galitzki, N., Golec,
  J.~E., Ho, S.-P.~P., Hargrave, P.~C., Kofman, A.~M., Lee, A.~T., Limon, M.,
  Matsuda, F.~T., Mauskopf, P.~D., Moodley, K., Nati, F., Niemack, M.~D.,
  Orlowski-Scherer, J., Page, L.~A., Partridge, B., Puglisi, G., Reichardt,
  C.~L., Sierra, C.~E., Simon, S.~M., Teply, G.~P., Tucker, C., Wollack, E.~J.,
  Xu, Z., and Zhu, N., ``The {S}imons {O}bservatory: modeling optical
  systematics in the {L}arge {A}perture {T}elescope,'' {\em Applied
  Optics}~{\bf 60},  823 (jan 2021).

\bibitem{Golec2020}
Golec, J.~E., McMahon, J.~J., Ali, A., Chesmore, G., Cooperrider, L., Dicker,
  S., Galitzki, N., Harrington, K., Jackson, R., Westbrook, B., Wollack, E.~J.,
  Xu, Z., and Zhu, N., ``Design and fabrication of metamaterial anti-reflection
  coatings for the {S}imons {O}bservatory,'' in [{\em Advances in Optical and
  Mechanical Technologies for Telescopes and Instrumentation
  {IV}}{\nolinebreak\hspace{0.1em}]},  Geyl, R. and Navarro, R., eds., {SPIE}
  (dec 2020).

\bibitem{Coughlin2018}
Coughlin, K.~P., McMahon, J.~J., Crowley, K.~T., Koopman, B.~J., Miller, K.~H.,
  Simon, S.~M., and Wollack, E.~J., ``Pushing the limits of broadband and
  high-frequency metamaterial silicon antireflection coatings,'' {\em Journal
  of Low Temperature Physics}~{\bf 193},  876--885 (May 2018).

\bibitem{datta2013}
Datta, R., Munson, C.~D., Niemack, M.~D., McMahon, J.~J., Britton, J., Wollack,
  E.~J., Beall, J., Devlin, M.~J., Fowler, J., Gallardo, P., Hubmayr, J.,
  Irwin, K., Newburgh, L., Nibarger, J.~P., Page, L., Quijada, M.~A., Schmitt,
  B.~L., Staggs, S.~T., Thornton, R., and Zhang, L., ``Large-aperture
  wide-bandwidth antireflection-coated silicon lenses for millimeter
  wavelengths,'' {\em Appl. Opt.}~{\bf 52},  8747--8758 (Dec 2013).

\bibitem{VavagiakisASC2021}
Vavagiakis, E., Ahmed, Z., Ali, A., Arnold, K., Austermann, J., Bruno, S.,
  Choi, S., Connors, J., Cothard, N., Dicker, S., Dober, B., Duff, S., Fanfani,
  V., Healy, E., Henderson, S., Ho, P., Hoang, D.-T., Hilton, G., Hubmayr, J.,
  and Zhu, N., ``The {S}imons {O}bservatory: {M}agnetic sensitivity
  measurements of microwave {SQUID} multiplexers,'' {\em IEEE Transactions on
  Applied Superconductivity}~{\bf PP},  1--1 (03 2021).

\bibitem{Crowley_2022}
Crowley, K.~D., Dow, P., Shroyer, J.~E., Groh, J.~C., Dober, B., Spisak, J.,
  Galitzki, N., Bhandarkar, T., Devlin, M.~J., Dicker, S., Gallardo, P.~A.,
  Harrington, K., Iuliano, J., Johnson, B.~R., Johnson, D., Kofman, A.~M.,
  Kusaka, A., Lee, A., Limon, M., Nati, F., Orlowski-Scherer, J., Page, L.,
  Randall, M., Teply, G., Tsan, T., Wollack, E.~J., Xu, Z., and Zhu, N., ``The
  {S}imons {O}bservatory: A large-diameter truss for a refracting telescope
  cooled to 1 {K},'' {\em Review of Scientific Instruments}~{\bf 93},  055106
  (May 2022).

\bibitem{dober_optical_2016}
Dober, B., Austermann, J.~A., Beall, J.~A., Becker, D., Che, G., Cho, H.~M.,
  Devlin, M., Duff, S.~M., Galitzki, N., Gao, J., Groppi, C., Hilton, G.~C.,
  Hubmayr, J., Irwin, K.~D., McKenney, C.~M., Li, D., Lourie, N., Mauskopf, P.,
  Vissers, M.~R., and Wang, Y., ``Optical {Demonstration} of {THz},
  {Dual}-{Polarization} {Sensitive} {Microwave} {Kinetic} {Inductance}
  {Detectors},'' {\em Journal of Low Temperature Physics}~{\bf 184},  173--179
  (July 2016).

\bibitem{Galitzki2016}
Galitzki, N., Ade, P., Angilè, F.~E., Ashton, P., Austermann, J., Billings,
  T., Che, G., Cho, H.-M., Davis, K., Devlin, M., Dicker, S., Dober, B.~J.,
  Fissel, L.~M., Fukui, Y., Gao, J., Gordon, S., Groppi, C.~E., Hillbrand, S.,
  Hilton, G.~C., Hubmayr, J., Irwin, K.~D., Klein, J., Li, D., Li, Z.-Y.,
  Lourie, N.~P., Lowe, I., Mani, H., Martin, P.~G., Mauskopf, P., McKenney, C.,
  Nati, F., Novak, G., Pascale, E., Pisano, G., Santos, F.~P., Scott, D.,
  Sinclair, A., Soler, J.~D., Tucker, C., Underhill, M., Vissers, M., and
  Williams, P., ``{Instrumental performance and results from testing of the
  BLAST-TNG receiver, submillimeter optics, and MKID detector arrays},'' in
  [{\em Millimeter, Submillimeter, and Far-Infrared Detectors and
  Instrumentation for Astronomy VIII}{\nolinebreak\hspace{0.1em}]},  Holland,
  W.~S. and Zmuidzinas, J., eds.,  {\bf 9914},  108 -- 118, International
  Society for Optics and Photonics, SPIE (2016).

\bibitem{austermann_millimeter-wave_2018}
Austermann, J.~E., Beall, J.~A., Bryan, S.~A., Dober, B., Gao, J., Hilton, G.,
  Hubmayr, J., Mauskopf, P., McKenney, C.~M., Simon, S.~M., Ullom, J.~N.,
  Vissers, M.~R., and Wilson, G.~W., ``Millimeter-{Wave} {Polarimeters} {Using}
  {Kinetic} {Inductance} {Detectors} for {TolTEC} and {Beyond},'' {\em Journal
  of Low Temperature Physics}~{\bf 193},  120--127 (Nov. 2018).

\bibitem{austermann_large_2018}
Austermann, J., Beall, J., Bryan, S.~A., Dober, B., Gao, J., Hilton, G.,
  Hubmayr, J., Mauskopf, P., McKenney, C., Simon, S.~M., Ullom, J., Vissers,
  M., and Wilson, G.~W., ``Large format arrays of kinetic inductance detectors
  for the {TolTEC} millimeter-wave imaging polarimeter ({Conference}
  {Presentation}),'' in [{\em Millimeter, {Submillimeter}, and {Far}-{Infrared}
  {Detectors} and {Instrumentation} for {Astronomy}
  {IX}}{\nolinebreak\hspace{0.1em}]},   {\bf 10708},  107080U, International
  Society for Optics and Photonics (July 2018).

\bibitem{AustermannSPIE22}
Austermann, J.~E., Beall, J.~A., Gao, J., Hubmayr, J., Ullom, J.~N., Vissers,
  M.~R., and Wheeler, J.~D., ``Aluminum-based millimeter-wave kinetic
  inductance detectors on 150~mm diameter substrates ({Conference} {Poster}),''
  in [{\em Millimeter, Submillimeter, and Far-Infrared Detectors and
  Instrumentation for Astronomy XI}{\nolinebreak\hspace{0.1em}]},  SPIE (July
  2022).

\bibitem{Sinclair2022}
Sinclair, A. et~al., ``{CCAT}-prime: {RFSoC} based readout for frequency
  multiplexed kinetic inductance detectors,'' {\em Paper No. 12190-66}  (2022).

\bibitem{VishwasSPIE22}
Vishwas, A., Parshley, S.~C., Gull, G.~E., Cortes-Medellin, G., Rossi, K.~M.,
  Herter, T.~L., Campbell, D.~B., Burnett, M.~C., Ammermon, S.~M., Ashcraft,
  N., Nygaard, E., Warnick, K.~F., Jeffs, B.~D., Mani, H., Groppi, C.~E., and
  Perillat, P., ``Alpaca: a fully cryogenic l-band phased array feed for radio
  astronomy ({Conference} {Presentation}),'' in [{\em Millimeter,
  Submillimeter, and Far-Infrared Detectors and Instrumentation for Astronomy
  XI}{\nolinebreak\hspace{0.1em}]},  SPIE (July 2022).

\bibitem{Koopman2020}
{Koopman}, B.~J., {Lashner}, J., {Saunders}, L.~J., {Hasselfield}, M.,
  {Bhandarkar}, T., {Bhimani}, S., {Choi}, S.~K., {Duell}, C.~J., {Galitzki},
  N., {Harrington}, K., {Hincks}, A.~D., {Ho}, S.-P.~P., {Newburgh}, L.,
  {Reichardt}, C.~L., {Seibert}, J., {Spisak}, J., {Westbrook}, B., {Xu}, Z.,
  and {Zhu}, N., ``{The Simons Observatory: overview of data acquisition,
  control, monitoring, and computer infrastructure},'' in [{\em Society of
  Photo-Optical Instrumentation Engineers (SPIE) Conference
  Series}{\nolinebreak\hspace{0.1em}]},  {\em Society of Photo-Optical
  Instrumentation Engineers (SPIE) Conference Series} {\bf 11452},  1145208
  (Dec. 2020).

\bibitem{Liu2017}
{Liu}, X., {Guo}, W., {Wang}, Y., {Wei}, L.~F., {Mckenney}, C.~M., {Dober}, B.,
  {Billings}, T., {Hubmayr}, J., {Ferreira}, L.~S., {Vissers}, M.~R., and
  {Gao}, J., ``Cryogenic led pixel-to-frequency mapper for kinetic inductance
  detector arrays,'' {\em Journal of Applied Physics}~{\bf 122},  034502 (July
  2017).

\end{thebibliography}
\bibliographystyle{spiebib} 

\end{document}